\documentclass{jps-cp}
\usepackage{txfonts} 

\title{
Weinberg Operator Contribution to the Hadronic CP Violation
}

\author{
Nodoka \textsc{Yamanaka}$^{1,2}$
}

\inst{$^{1}$Kobayashi-Maskawa Institute for the Origin of Particles and the Universe, Nagoya University, Furocho, Chikusa, Aichi 464-8602, Japan\\
$^{2}$
Nishina Center for Accelerator-Based Science, RIKEN, Wako 351-0198, Japan
}

\email{
nyamanaka@kmi.nagoya-u.ac.jp
}

\recdate{December 13, 2021}

\abst{
The CP violating Weinberg operator (or the gluon chromo-electric dipole moment) is generated in many candidates of new physics beyond the standard model, and it contributes to observables such as the electric dipole moments (EDM) of the neutron, nuclei and atoms.
In this proceedings contribution, we review the present understanding of the Weinberg operator as well as the result of the estimation of its effect to the EDM of $^3$He nucleus through the contact CP-odd nuclear force.
}

\kword{
CP violation, electric dipole moment, hadron
}

\begin{document}
\maketitle

\section{Introduction}

It is believed that the baryon number asymmetry which may be observed today was generated in the early stage of the Universe.
However, one of the required conditions, CP violation in fundamental theory \cite{Sakharov:1967dj}, is lacking in the standard model (SM) \cite{Kobayashi:1973fv,Huet:1994jb}.
We therefore need a new theory beyond the SM with large CP violating interactions, and these may be probed using the electric dipole moment (EDM), which is measurable in many systems \cite{Yamanaka:2014mda,Yamanaka:2017mef,Chupp:2017rkp}.

The EDMs of composite systems such as the nucleon, nuclei, or atoms, are not generated by the EDM of elementary particles which compose them, but also by CP violating interactions between them.
An important example of such interactions is the {\it Weinberg operator} \cite{Weinberg:1989dx,Bigi:1990kz,Bigi:1991rh}, defined as
\begin{eqnarray}
{\cal L}_w 
&=& 
\frac{1}{3!} w 
f^{abc} \epsilon^{\alpha \beta \gamma \delta} G^a_{\mu \alpha } G_{\beta \gamma}^b G_{\delta}^{\ \ \mu,c}
,
\label{eq:weinberg_operator}
\end{eqnarray}
and it appears in many candidates such as the extension of the SM Higgs sector, while it is very small in the SM \cite{Pospelov:1994uf,Yamaguchi:2020dsy}.

The EDMs of the neutron, $^{199}$Hg and $^{129}$Xe atoms were recently measured in several experiments, with the upper limits \cite{Abel:2020gbr,Graner:2016ses,Sachdeva:2019rkt}
\begin{eqnarray}
|d_n| 
& < & 
1.8 \times 10^{-26} e \, {\rm cm}
,
\\
|d_{\rm Hg}| 
& < & 
7.4 \times 10^{-30} e \, {\rm cm}
,
\\
|d_{\rm Xe}| 
&<& 
1.4 \times 10^{-27} e \, {\rm cm}
.
\end{eqnarray}
We see that the EDMs are very tightly bound, and we expect the Weinberg operator coupling $w$ and the new physics behind it to be strongly constrained.
The relation between the neutron EDM and $w$, while still being difficult to quantify in lattice \cite{Dragos:2019oxn,Rizik:2020naq}, has been investigated using many other approaches \cite{Demir:2002gg,Haisch:2019bml,Hatta:2020ltd,Yamanaka:2020kjo,Hatta:2020riw,Weiss:2021kpt}.
On the other hand, the effect of $w$ to the CP-odd nuclear force, which presumably gives the leading contribution to the atomic and nuclear EDMs, has never been discussed in detail.

In this proceedings contribution, we review the current understanding of the hadron level contribution of the Weinberg operator and its effect on observable EDMs.
In the next section, we derive the hadron level effective field theory of the Weinberg operator including the nucleon EDM.
In section \ref{sec:nuclear}, we estimate the contribution of the contact CP-odd nuclear force generated by the Weinberg operator in the EDM of $^3$He nucleus.
The final section is devoted to the summary.

\section{Hadron level effective field theory of the Weinberg operator}

\begin{figure*}[t]
\begin{center}
\includegraphics[width=0.5\textwidth]{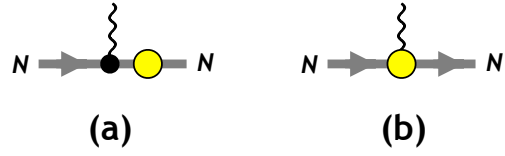}
\caption{Schematic picture of the nucleon EDM generated by chiral rotation (a) and by the irreducible process (b).}
\label{fig:NEDM}
\end{center}
\end{figure*}

In the lowest order of hadron effective field theory, the nucleon EDM has two leading contributions, namely the chiral rotation of the nucleon anomalous magnetic moment (g-2) [Fig. \ref{fig:NEDM} (a)], and the irreducible (contact) diagram [Fig. \ref{fig:NEDM} (b)] \cite{Bigi:1990kz,Bigi:1991rh}.
In recent work, it was claimed that the first one gives the dominant contribution \cite{Demir:2002gg,Haisch:2019bml}.
Indeed, by redefining the nucleon mass Lagrangian as
\begin{eqnarray}
-m_N \bar N N
-m_{CP} \bar N i \gamma_5 N
\to
-m_{N'} \bar N' N'
,
\end{eqnarray}
the nucleon g-2 $\mu_N$ outputs an EDM as
\begin{eqnarray}
\mu_{N}
\bar N \sigma_{\mu \nu} F^{\mu \nu} N
\to
\mu_{N'}
\bar N' \sigma_{\mu \nu} F^{\mu \nu} N'
+
d_{N'}
\bar N' i \sigma_{\mu \nu} \gamma_5 F^{\mu \nu} N'
.
\end{eqnarray}
When $m_{CP}\ll m_N$, we have $m_{N'} \approx m_N$, and the nucleon EDM is 
\begin{eqnarray}
d_{N'}
\approx
\frac{m_{CP}}{m_N} \mu_N
.
\end{eqnarray}
The CP-odd nucleon mass generated by the Weinberg operator has been calculated using QCD sum rules \cite{Demir:2002gg,Haisch:2019bml}:
\begin{equation}
m_{CP}
=
-
\langle N | 
{\cal L}_w 
| N \rangle
=
-m_N w \frac{3 g_s m_0^2}{32 \pi^2}
\ln \left( \frac{M^2}{\mu_{\rm IR}^2} \right)
,
\end{equation}
with $m_0^2 = (0.8 \pm 0.2 )$ GeV$^2$, $g_s = 2.13 \pm 0.03$, and $\frac{M}{\mu_{\rm IR}} = \sqrt{2} (1.5 \pm 0.5)$ \cite{Haisch:2019bml}.
The nucleon EDM induced by the chiral rotation is then
\begin{eqnarray}
d_N^{\rm (a)} (w) 
&\approx&
w \times 
\left\{
\begin{array}{rl}
 (25\pm 13) \, e \, {\rm MeV} & (N = n ) \cr
(-23\pm 12) \, e \, {\rm MeV} & (N = p ) \cr
\end{array}
\right.
.
\label{eq:weinbergop_red}
\end{eqnarray}
The above result, however, does not contain the irreducible nucleon EDM [see Fig. \ref{fig:NEDM} (b)].
This contribution may actually be calculated in the nonrelativistic quark model using the Gaussian expansion method \cite{Hiyama:2003cu} by using the same techniques as those in the evaluation of the nuclear EDM  \cite{Yamanaka:2015qfa,Yamanaka:2015ncb,Yamanaka:2016itb,Yamanaka:2016fjj,Yamanaka:2016umw,Lee:2018flm,Yamanaka:2019vec}.
The result is \cite{Yamanaka:2020kjo}
\begin{eqnarray}
d_N^{\rm (b)} (w) 
& = &
w \times 
\left\{
\begin{array}{rl}
- (4 -5) \, e \, {\rm MeV} & (N = n ) \cr
(4 -5) \, e \, {\rm MeV} & (N = p ) \cr
\end{array}
\right.
.
\label{eq:weinbergop_irred}
\end{eqnarray}
We see that the irreducible contribution is smaller than that generated by the chiral rotation, which confirms the assumption made in earlier works.
By combining both results, we have 
\begin{eqnarray}
d_N (w) 
& = &
w \times 
\left\{
\begin{array}{rl}
(20 \pm 12) \, e \, {\rm MeV} & (N = n ) \cr
-(18 \pm 11) \, e \, {\rm MeV} & (N = p ) \cr
\end{array}
\right.
,
\label{eq:weinbergop_edm}
\end{eqnarray}
where the theoretical uncertainty was estimated by assuming that $d_N^{\rm (b)} (w)$ has an $O(100\%)$ systematic error.

\section{Nuclear level CP violation\label{sec:nuclear}}

\begin{figure*}[t]
\begin{center}
\includegraphics[width=0.5\textwidth]{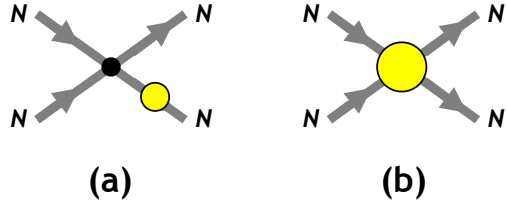}
\caption{Schematic picture of the CP-odd nuclear force generated by the chiral rotation (a) and by the irreducible process (b).}
\label{fig:CPVNN}
\end{center}
\end{figure*}

Let us now apply the same analysis to the CP-odd nuclear force.
Since the Weinberg operator has no flavor dependence, it is in principle possible to match it with the isoscalar CP-odd nuclear force.
We note however that the Weinberg operator does not contribute to the isoscalar one-pion exchange CP-odd nuclear force at the leading order of chiral perturbation \cite{deVries:2011an,Yamanaka:2016umw}.
This is because the isoscalar CP-odd pion-nucleon interaction
\begin{equation}
{\cal L}_{\pi NN}^{(0)}
=
\bar g_{\pi NN}^{(0)}
\sum_{a=1}^3
\bar N \tau_a N
\pi_a
,
\end{equation}
is chiral symmetry breaking, while the Weinberg operator Lagrangian (\ref{eq:weinberg_operator}) is chiral symmetry conserving (purely gluonic).
This implies that $\bar g_{\pi NN}^{(0)}$ generated by $w$ is at least suppressed by a factor of light quark mass.
The leading contribution is then given by the CP-odd contact interaction
\begin{equation}
{\cal L}_C
=
-\bar C_1
m_N
\bar N N \, 
\bar N i \gamma_5 N
-
\bar C_2
m_N
\sum_{a=1}^3
\bar N \tau_a N \, 
\bar N i \gamma_5 \tau_a N
,
\label{eq:contactCPVNN}
\end{equation}
which has a chiral rotated part, calculable in a similar manner as the nucleon EDM, and an irreducible part (see Fig. \ref{fig:CPVNN}).
The nuclear force of Eq. (\ref{eq:contactCPVNN}) has a delta function-like shape in the coordinate space.
By using the limiting formula of the Yukawa function $\lim_{m \to \infty} m^2 e^{-m r} / 4 \pi r= \delta^{(3)} (\vec{r})$, we can match Eq. (\ref{eq:contactCPVNN}) with the following vector meson exchange CP-odd nuclear force
\begin{equation}
V_{CPVNN} (\vec{r})
=
\frac{-1}{8 \pi m_N}
(\vec{\sigma}_1 - \vec{\sigma}_2) \cdot 
\Biggl[
\bar G_\rho^{(0)} 
\sum_{a=1}^3 (\tau_{1a} \tau_{2a})
\vec{\nabla } \frac{e^{-m_\rho r}}{r} 
+\bar G_\omega^{(0)} 
\vec{\nabla } \frac{e^{-m_\omega r}}{r} 
\Biggr]
,
\label{eq:vectormesonCPVNN}
\end{equation}
where $\tau_{i}$ and $\vec{\sigma}_i$ denote the isospin and spin of the $i$-th interacting nucleon, and $\vec{r}$ is the relative coordinate between the two nucleons, directed to the nucleon 1 (the direction of $\vec{\nabla}$ also follows this convention).
The masses of the vector mesons are $m_\rho = 770$ MeV and $m_\omega = 780$ MeV, but they are not important in the derivation as long as they are considered to be heavy.
The matching of the couplings is then
\begin{eqnarray}
\bar G_\omega^{(0)}
&\approx&
m_N m_\omega^2 \bar C_1
,
\\
\bar G_\rho^{(0)}
&\approx&
m_N m_\rho^2 \bar C_2
.
\end{eqnarray}
Since the light nuclei's EDMs were calculated using the CP-odd vector meson exchange potential (\ref{eq:vectormesonCPVNN}) \cite{Song:2012yh,Yamanaka:2015qfa,Yamanaka:2016umw,Froese:2021civ}, it is possible to derive the dependence of the EDMs of these nuclei on $w$ if we know the relationship between the contact couplings $\bar C_1$, $\bar C_2$ and $w$.
Here, $\bar C_1$ and $\bar C_2$ are derived from the Weinberg operator and the nucleon EDM is estimated by chirally rotating the CP-even contact nuclear force.
The CP-even contact interaction
\begin{equation}
V_{NN}
=
\frac{1}{2} C_S
\bar N N \, 
\bar N N
,
\label{eq:NN}
\end{equation}
was fitted in the chiral perturbation theory with a cutoff around 1 GeV.
From Ref. \cite{Epelbaum:2008ga}, we have $C_S = -120.8$ GeV$^{-2}$, while Refs. \cite{Weinberg:1991um,Bernard:1995dp} estimated $|C_S| \approx 290$ GeV$^{-2}$ from the deuteron binding energy, where we assumed that $C_S$ gives the dominant contribution.
The chiral rotation of Eq. (\ref{eq:NN}) simply yields
\begin{equation}
m_N \bar C_1
=
\frac{m_{CP}}{m_N} C_S
=
\frac{\bar G_\omega^{(0)}}{ m_\omega^2 }
.
\end{equation}

Let us give a concrete value for the EDM of $^3$He nucleus generated by the Weinberg operator via the CP-odd nuclear force.
The nuclear level calculation of the direct contribution of $\bar C_1$ was performed in chiral perturbation theory \cite{Bsaisou:2014oka}, giving
\begin{equation}
d_{\rm He} [ \bar C_1 (w)]
=
(0.04 \pm 0.02)
\bar C_1 (w)
e\, {\rm fm}^{-2}
=
(25 \pm 12)
w\,
e\, {\rm MeV}
,
\label{eq:contactresultEFT}
\end{equation}
while we have in the vector meson exchange model \cite{Song:2012yh,Yamanaka:2015qfa,Yamanaka:2016umw,Froese:2021civ}
\begin{equation}
d_{\rm He} [ \bar G_\omega^{(0)} (w)]
=
[2 -8] \times 
10^{-4}
\bar G_\omega^{(0)} (w)
e\, {\rm fm}
=
[2-7]
w\,
e\, {\rm MeV}
,
\label{eq:contactresultmeson}
\end{equation}
where the range denotes the systematic uncertainty due to the choice of the realistic CP-even nuclear force.
Here we used $C_S$ of Ref. \cite{Epelbaum:2008ga}, but we have not considered the uncertainty related to the fit.
We should consider the difference between the numbers in Eqs. (\ref{eq:contactresultEFT}) and (\ref{eq:contactresultmeson}) as the systematic uncertainty of nuclear physics.

On the other hand, the intrinsic nucleon EDM contribution to $^3$He EDM is
\begin{equation}
d_{\rm He} [d_N (w)]
=
-0.04 \, d_p (w) + 0.9 \, d_n (w)
=
(19 \pm 11)
w\,
e\, {\rm MeV}
,
\end{equation}
where we used the values of Eq. (\ref{eq:weinbergop_edm}).
We see that the contribution from the contact CP-odd nuclear force (\ref{eq:contactCPVNN}) and that of the intrinsic nucleon EDM are comparable.
Here we have neglected the effect of the irreducible diagram [Fig. \ref{fig:CPVNN} (b)], but we anticipate that it is not sizable according to the analysis of the nucleon EDM.

\section{Summary}

In this proceedings contribution, we discussed the nucleon and nuclear EDMs generated by the Weinberg operator.
Due to the chiral symmetry, the Weinberg operator does not generate the one-pion exchange CP-odd nuclear force, and the polarization is only induced by the contact interaction, with a contribution comparable to that of the valence nucleon, at least in the case of $^3$He.

In this analysis, we neglected the irreducible contribution to the CP-odd nuclear force, but this has to be evaluated in the future to confirm whether its effect is really small or not.
We expect that we can also calculate the quark model as in the case of the nucleon EDM.
We also note that large nuclear EDM and the Schiff moment \cite{Yanase:2020agg} may be generated by the Weinberg operator if the long range pion-exchange nuclear force is induced by the vacuum alignment mechanism.
These evaluations are left for future research.

Note added: the long-range pion-exchange nuclear force induced by the Weinberg operator was recently calculated in Ref. \cite{Osamura:2022rak} and it was found that this contribution is not negligible for the atomic EDM.

\end{document}